# Parity-Time Synthetic Laser

Liang Feng[1†], Zi Jing Wong[1†], Renmin Ma[1†], Yuan Wang[1,2] and Xiang Zhang[1,2]*

[1]*NSF Nanoscale Science and Engineering Center, 3112 Etcheverry Hall, University of California, Berkeley, California 94720, USA*

[2]*Materials Sciences Division, Lawrence Berkeley National Laboratories, 1 Cyclotron Road, Berkeley, California 94720, USA*

[†]These authors contributed equally to this work.

*E-mail: xiang@berkeley.edu

Parity-time (PT) symmetry is a fundamental notion in quantum field theories[1,2]. It has opened a new paradigm for non-Hermitian Hamiltonians ranging from quantum mechanics, electronics, to optics. In the realm of optics, optical loss is responsible for power dissipation, therefore typically degrading device performance such as attenuation of a laser beam. By carefully exploiting optical loss in the complex dielectric permittivity, however, recent exploration of PT symmetry revolutionizes our understandings in fundamental physics and intriguing optical phenomena such as exceptional points and phase transition that are critical for high-speed optical modulators[3-9]. The interplay between optical gain and loss in photonic PT synthetic matters offers a new criterion of positively utilizing loss to efficiently manipulate gain and its associated optical properties[10-19]. Instead of simply compensating optical loss in conventional lasers, for example, it is theoretically proposed that judiciously designed delicate modulation of optical loss and gain can lead to PT synthetic lasing[20-23] that fundamentally broadens laser physics. Here, we report the first experimental demonstration of PT synthetic lasers. By carefully exploiting the interplay between gain and loss, we achieve degenerate eigen modes at the same frequency but with complex conjugate gain and loss coefficients. In contrast to conventional ring cavity lasers with multiple modes, the PT synthetic micro-ring laser exhibits an intrinsic single mode lasing: the non-threshold PT broken phase inherently associated in such a photonic system enables a single lasing mode regardless of the gain spectral bandwidth. This chip-scale semiconductor platform provides a unique route towards fundamental explorations of PT physics and next generation of optoelectronic devices for optical communications and computing.

The mathematical concept of PT symmetry was firstly introduced in quantum mechanical systems more than a decade ago. However, the stringent requirement for the sophisticated interplays of complex perturbations in Hamiltonians is difficult to realize PT symmetry experimentally in quantum mechanical systems. Through the quantum-optical analogue, optics with delicately balanced gain and loss has become a realistic platform to study the fundamentals of PT symmetry and associated unique properties such as unidirectional light transport. Although loss is usually undesirable in conventional optical devices such as optical emitters and lasers[24-28], its delicate interplay with gain can counter-intuitively realize unprecedented lasing cavities. Additionally, judicious loss manipulation may also enable one to efficiently control directional lasing and light amplification[3]. Therefore, PT synthetic lasing is not only important to fundamental understandings of PT symmetry, but also opens a new realm in optical physics and laser applications.

In this letter, we experimentally demonstrate the first PT synthetic lasers by exploring the PT broken phase to achieve simultaneously the energy-degenerate lasing and absorption modes. We devised the PT-modulated synthetic micro-ring lasers on a chip-scale III-V semiconductor platform. Our design, for the first time, realized the non-threshold PT symmetry breaking taking the advantage of the continuous rotational symmetry for the desired whispering-gallery-mode (WGM) order in the micro-ring resonator. Because of such a rotational symmetry, two energy-degenerate modes with complex conjugate modal gain/loss coefficients coexist in the PT synthetic laser, leading to the PT symmetry breaking. However, all other WGM orders cannot "see" the desired rotational symmetry of modulation and the corresponding fields are uniformly distributed in gain and loss regions, resulting in no net modal amplification or absorption and thus falling into PT symmetric phase. As a result, no matter how large the optical gain spectrum is, only the lasing mode at the desired WGM order overlap precisely with the gain regions in the ring with a minimum effect from the introduced loss and therefore contain sufficient modal gain above the lasing threshold; whereas all other WGM modes damp away by the intentionally introduced loss in PT-modulation, leading to an intrinsic single-mode WGM lasing. In contrast to multi-mode lasing from typical ring resonators, the demonstrated single-mode WGM lasing is only associated with the specific WGM order satisfying the PT broken phase, which is an intrinsic broadband characteristics regardless of the gain spectral bandwidth.

As depicted in Fig. 1a, the PT synthetic micro-ring resonator is designed with 500 nm-thick InGaAsP multiple quantum wells (MQWs) on an InP substrate. InGaAsP MQWs have a high material gain coefficient over 1,000 cm$^{-1}$ around 1,500 nm[29] sufficient for PT synthetic lasing. The PT synthetic gain and loss modulation is periodically introduced using additional Gemanium (Ge)/Chrome (Cr) structures on top of the InGaAsP MQW along the azimuthal direction:

$$\Delta n = \begin{cases} n_{gain} = -in" & (l\pi/m < \varphi < (l+1/2)\pi/m) \\ n_{loss} = in" & ((l+1/2)\pi/m < \varphi < (l+1)\pi/m) \end{cases},$$

where $n"$ denotes the index modulation in only the imaginary part, and $m$ the azimuthal order of the desired WGM mode in the micro-ring resonator and $l = 0, 1, 2 ... 2m-1$ divides the micro-ring into $2m$ periods. The thickness of Ge/Cr structures is carefully designed to introduce loss that exactly reverses the sign of the imaginary part of the local modal index, while maintaining the same real part. Using the coupled mode theory, the wavenumbers of the eigen WGM modes in the PT micro-ring resonator can be designed as $\beta = \beta_0 \pm i\kappa n"$ (see Methods)[30], where $\beta$ is the complex wavenumber of the WGM mode in such a PT structure, $\beta_0$ is the intrinsic wavenumber of the WGM mode without gain and loss, and $\kappa$ denotes the coupling between the clockwise and counter-clockwise traveling waves through the periodic PT modulation in the micro-ring resonator. We intentionally work on the mode order of $m = 53$, owing to its WGM wavelength is centered at the gain spectrum of InGaAsP MQWs. Evolution of PT symmetry in the synthetic micro-ring resonator can be seen from the corresponding complex eigen spectra (Figs. 1b and 1c). Two WGM modes are energy-degenerate at the same eigen resonant frequency ($f = c\,\text{Re}(\beta)/2\pi$), but their modal wavenumbers are complex conjugated with each other, corresponding to a PT broken phase. In other words, in the PT broken phase, two WGM modes at the same eigen frequency coincide in the micro-ring: one lasing mode associated with net gain but another absorption mode with loss, realizing simultaneous coexistence of lasing and anti-lasing eigen modes. It is worth noting that although effective modal gain is obtained for the lasing mode, energy cannot be accumulated to infinity in the micro-ring resonator as light has to escape from the resonator with a finite quality factor to reach the equilibrium state when above the lasing threshold. More interestingly, the PT phase transition occurs immediately and the micro-ring resonator goes into the PT broken phase even if the strength of gain-loss modulation is infinitesimal. The observed non-threshold feature in our PT phase transition in the synthetic micro-ring resonator is owing to the continuous rotational symmetry associated with the desired WGM order in absence of real index modulation. If the real index modulation is introduced, phase transition from the PT symmetric to broken phase can be expected (see Supplementary Information). On the other hand, the observed non-threshold PT symmetry breaking is unique compared to the previously studied finite coupled waveguide systems in which there are no continuous symmetries and PT modulation is always truncated, such that PT symmetry breaking must require finite strength of gain-loss modulation in spite of absence of real index modulation[23].

Figs. 1d and 1e show the modal intensity distribution of the lasing and absorption WGM modes in the PT micro-ring resonator, respectively: one with electric fields strongly confined in the amplification sections exhibiting net modal gain and thus above the lasing threshold, while the other is spatially separated $\pi/2$ in phase with electric fields mainly under the Cr/Ge sections therefore becoming loss dominant. Additionally, their eigen frequencies are energy-degenerate but their wavenumbers are complex conjugates with each other, consistent with the features of the PT broken phase. For other WGM modes, for instance $m = 54$, electric fields in both energy-degenerate WGM modes are uniformly distributed in gain and loss sections, creating almost real modal wavenumbers without net gain and loss (see Supplementary Information). In other words, the continuous rotational symmetry of the introduced PT-modulation cannot be precisely matched by undesired WGM orders. Because of balanced gain and loss in the PT synthetic resonator, the corresponding undesired WGM modes (such as $m=54$) are similar to modes in a real-indexed micro-ring resonator, corresponding the PT symmetric phase without modal gain or loss. It is evident that the introduced loss minimally affects the lasing mode at the desired WGM order, but intensively suppress all other modes no matter how large the gain spectra are, strongly squeezing optical gain of the whole spectrum into the desired WGM order and thus leading to an intrinsic single-mode WGM lasing. This unique single-mode operation is even valid with an arbitrarily wide gain spectrum, inherently different from the conventional single-mode micro-ring lasers that use the real-index coupling and modulation to achieve the mode splitting in eigen frequencies which is limited by the bandwidth of gain media[31].

The MQW based PT ring laser with Cr/Ge modulations are fabricated by Ebeam lithography (Fig. 2a). Under optical pumping, broad photoluminescence (PL) emission around 1,500 nm was first observed at low pump power densities (Fig 2b). As pump power was increased, the amplified spontaneous emission (ASE) spectrum showed a drastic spectral narrowing. Although below the lasing threshold, the single WGM resonance of the cavity mode clearly emerged. At higher pumping intensities well above the lasing threshold, a single-mode lasing peak occurred at the wavelength of 1,513 nm, confirming our theoretical prediction of single WGM mode lasing oscillation from the PT synthetic laser. In Fig. 2c, the light-light curve corresponding to the single-mode lasing emission clearly shows the slope change, an indication of the lasing threshold at pump power density of about 600 MW cm$^{-2}$. While the desired mode ($m=53$) in the PT broken phase emerges above the lasing threshold, creating well-pronounced single-mode WGM lasing with an extinction ratio of more than 20 dB, other WGM modes including the absorption mode at the same azimuthal order are all below the lasing threshold and damp away (for example, $m=54$ in Supplementary Information).

For comparison, a control sample of a WGM laser was fabricated consisting of the same-sized InGaAsP/InP micro-ring resonator without additional Cr/Ge periodic index modulation azimuthally. As expected, we observed a typical multimode lasing spectrum with different WGM azimuthal orders distributed over the gain spectral region (Fig. 3a). Compared to the PT synthetic laser, it can be seen that at the same pumping condition, the resonance peak for the same azimuthal order of $m=53$ well matches the single-mode lasing of PT ring resonator at the wavelength of 1,513 nm (Fig. 3b). Moreover, we also fabricated an additional PT synthetic laser with the same size but different azimuthal PT modulation for the order of $m=55$. Its lasing emission at 1,467 nm (Fig. 3b) also agrees well with the multi-mode lasing spectrum of the conventional WGM laser for the same azimuthal order (Fig. 3a). The slight deviation in resonance wavelengths maybe resulted from the varying material properties of InGaAsP MQWs at different wavelengths including the index of refraction and the material gain coefficient. It is evident that instead of altering the WGM mode in the micro-ring resonator, the introduced PT gain/loss modulation selects the lasing WGM mode in the PT broken phase over a broad band of spectrum. By changing the desired azimuthal order of the structured PT modulation, the single-mode lasing frequency can be efficiently tuned. We may also envision exploring PT synthetic lasers at a nanoscale beyond the diffraction limit with strong light-matter interactions on plasmonic platforms[32,33].

We have demonstrated the first PT synthetic laser by the delicate exploitation of optical loss and gain, in such a way that only the desired WGM order becomes amplification and absorption degenerated in the PT broken phase, while others stay gain-loss balanced in the PT symmetric phase. Such a PT synthetic laser is intrinsically single-mode, in stark contrast to typical multi-mode lasing in micro-ring resonator. We showed that non-threshold PT broken phase enables single-mode lasing. More importantly, the chip-scale semiconductor demonstration not only offers an exciting experimental platform to explore fundamental PT symmetric quantum mechanics[34] but also is a major step towards unique photonic devices such as a PT symmetric laser-absorber that coincides lasing and anti-lasing (i.e. coherent perfect absorption[35,36]) simultaneously.

**Methods**
**Derivation of eigen wavenumbers of the WGM modes in the micro-ring resonator.**
Assuming a small variation and negligible scattering loss in the PT synthetic micro-ring resonator, the coupled mode equations can be written as

$$\begin{cases} \dfrac{dA}{Rd\theta} = i\beta_0 A + in''\kappa B \\ \dfrac{dB}{Rd\theta} = i\beta_0 B - in''\kappa A \end{cases},$$

where A and B denote the clockwise and counter-clockwise waves traveling in a harmonic form of $A = A_0 e^{i\beta R\theta}$ and $B = B_0 e^{i\beta R\theta}$, respectively, $R$ is the radius of the micro-ring resonator, and $\theta$ is the corresponding azimuthal angle. By solving the coupled mode equations, the eigen wavenumbers of the WGM modes supported in the micro-ring resonator can be obtained,

$$\beta = \beta_0 \pm i\kappa n'',$$

manifesting non-threshold PT symmetry breaking in our PT synthetic lasers.

**Fabrication and characterization of PT synthetic lasers.** The PT synthetic laser was fabricated using CMOS-compatible overlay electron beam lithography with accurate alignment, followed by evaporation of Cr/Ge as well as dry etching to form the InGaAsP/InP micro-resonator. The fabricated PT synthetic laser was optically pumped: a Ti:Sapphire pulsed laser was employed operating at the repetition rate of 80 MHz and with a pulse width of 100 fs. The pump beam was delivered to the device using a ×20 long-working distance objective that was also used to collect the light emission for the lasing spectral analysis. The emitted light was coupled into a spectrometer for spectral analysis.

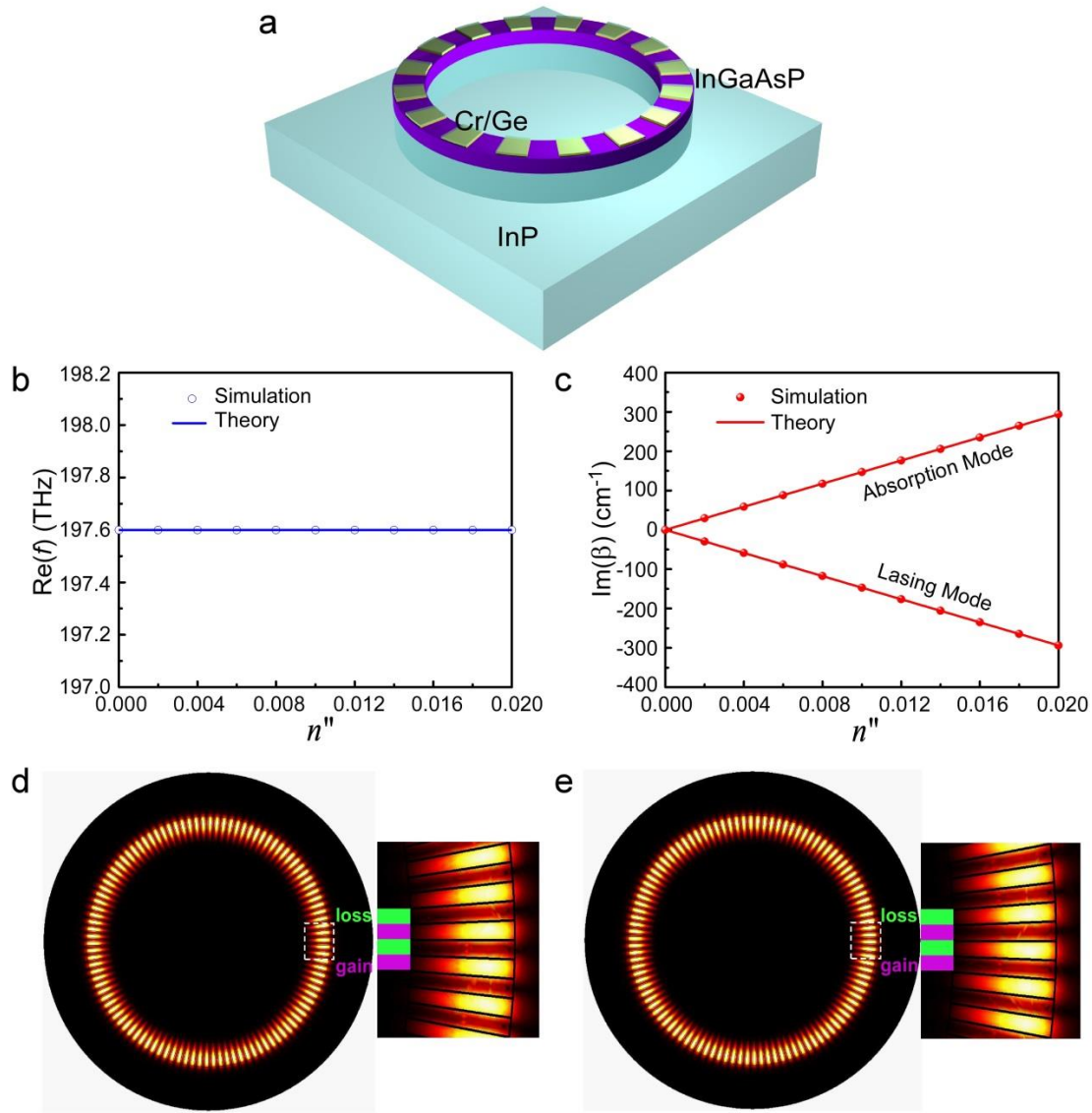

**Figure 1. Design of PT synthetic lasers. a,** Schematic of the parity-time synthetic laser consists of Cr/Ge bilayer structures periodically arranged in the azimuthal direction on top of the InGaAsP/InP micro-ring resonator to mimic a pure gain-loss modulation. The diameter of the micro-ring resonator is set to be 8.9 μm in order to achieve a high quality-factor, thus lowering the corresponding lasing threshold. To improve the Q-factor, we extend the micro-ring 1 μm-deeper into the InP substrate from the top InGaAsP MQW layer. In order to avoid mode competition, the width of the micro-ring resonator is chosen to be 900 nm, such that only single-radial-order WGM modes are supported. Here, the designed azimuthal order of the modulation is $m = 53$, in order to achieve the resonant WGM wavelength around 1,500 nm. **b** shows the eigen frequency of the WGM modes

for $m = 53$ and **c** is the complex conjugated imaginary eigen spectra at the same azimuthal order by numerical simulations (circles/dots) and theoretical calculations (solid lines), indicating a non-threshold PT symmetry breaking. In the PT broken phase, an energy-degenerate pair of complex conjugated WGM modes occurs: one for lasing and the other for absorption, which is the necessary condition to achieve PT synthetic lasers. In **d** and **e**, the PT modulation is designed with 15nm Cr/40nm Ge, corresponding to gain-loss modulation at a coefficient of 800 cm$^{-1}$. Eigen electric field distribution of paired lasing and absorption WGM modes can be observed with fields confined in the gain (**d**) and loss (**e**) sections, respectively. They share the same eigen frequency of 197.6 THz, but possess conjugate modal gain/loss coefficients with $\text{Im}(\beta) = -134$ cm$^{-1}$ for the lasing mode (**d**) and $\text{Im}(\beta) = 134$ cm$^{-1}$ for the absorption mode (**e**), respectively.

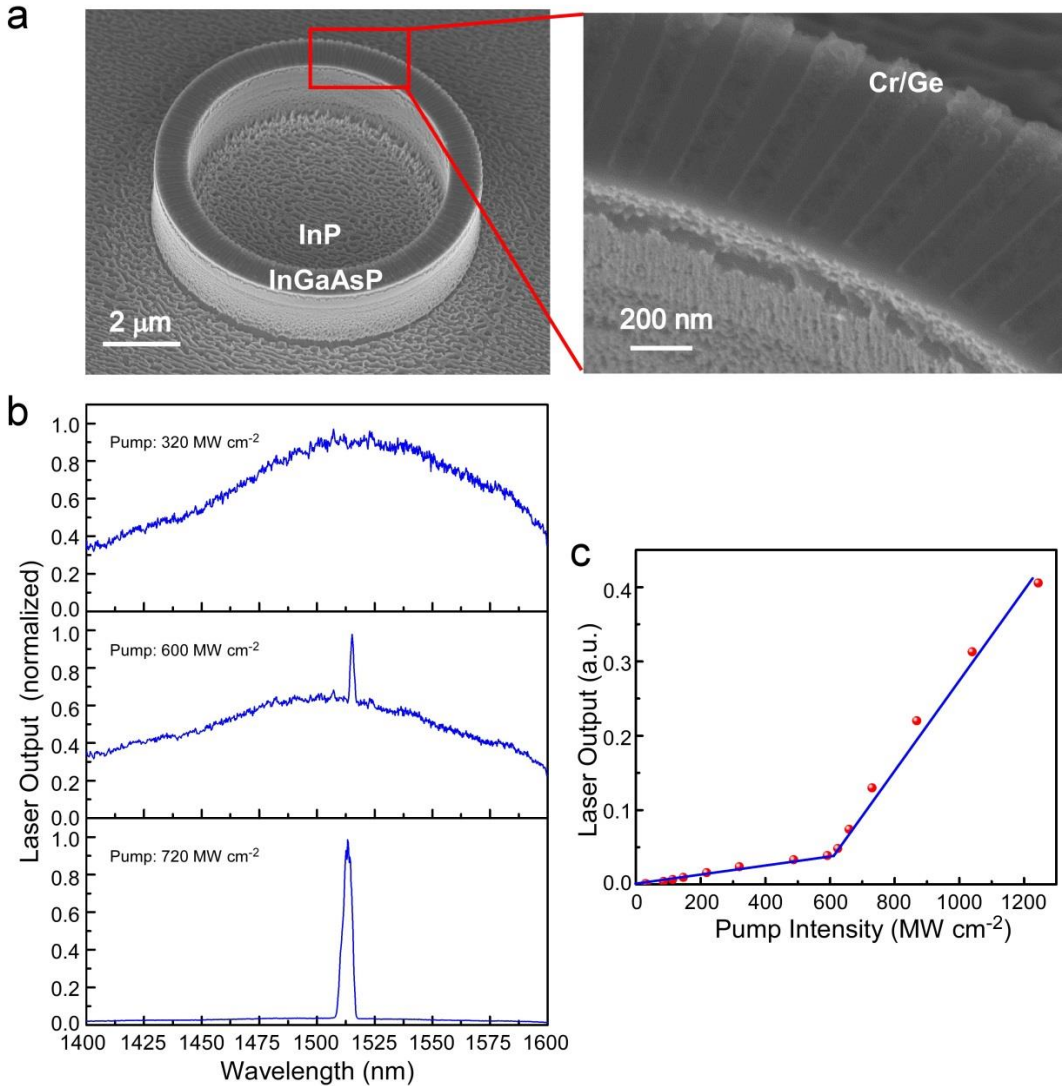

**Figure 2. Experimental characterization of PT synthetic laser. a,** Scanning electron microscope images of the fabricated PT synthetic laser. The Cr/Ge structures are periodically arranged on top of the InGaAsP/InP micro-ring in the azimuthal direction to create the pure balanced gain/loss modulation for PT synthetic lasing. **b,** Evolution of light emission spectra from PL, ASE, to single-mode WGM lasing at the wavelength of about 1,513 nm as the intensity of pump light was increased from 320 MW cm$^{-2}$, 600 MW cm$^{-2}$, to 720 MW cm$^{-2}$, respectively. At 320 MW cm$^{-2}$, it can be seen that the gain spectrum is centered around 1,510 nm where our desired WGM order is designed for. At 600 MW cm$^{-2}$, ASE can be seen as the WGM resonance appears in the light emission spectrum. Finally, at 720 MW cm$^{-2}$, the broadband single-mode lasing from our PT

synthetic laser is convincingly demonstrated with a high extinction ratio of more than 20 dB. **c,** Light-light curve of the PT synthetic laser shows the power relationship between the lasing emission and the pump light, convincingly validating the onset of the intrinsic single-mode lasing at the threshold slightly larger than 600 MW cm$^{-2}$ in our PT synthetic laser.

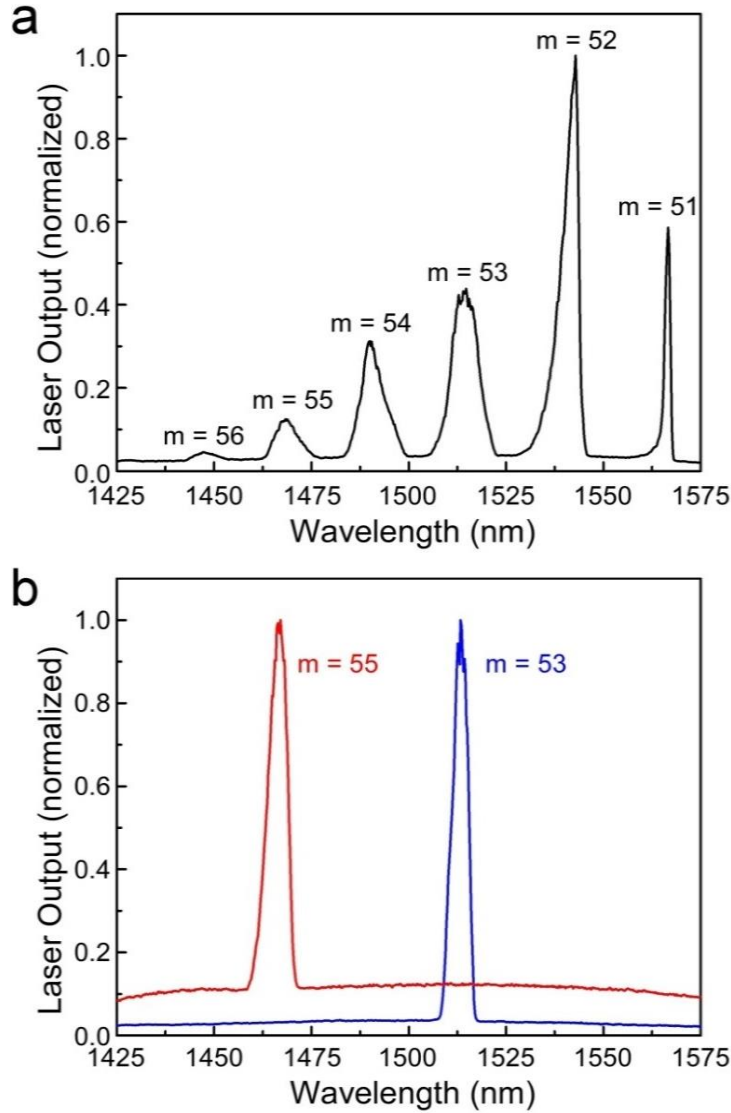

**Figure 3. Comparison between PT synthetic laser and typical micro-ring WGM laser. a,** Typical multi-mode lasing spectrum observed from the micro-ring WGM laser without any index modulation, showing a series of lasing modes corresponding to different azimuthal orders. These azimuthal orders are confirmed using numerical simulations. The lasing modes are highly distributed in the whole spectrum since all the modes share limited optical gain to achieve their respective lasing thresholds. **b,** Single-mode lasing spectra of the PT synthetic lasers corresponding to the azimuthal orders of $m=53$ and $m=55$, consistent with the lasing wavelength for the same azimuthal orders in **a**. This confirms that our PT synthetic laser does not alter original WGM modes but efficiently select the desired lasing mode, squeezing optical gain from the whole spectrum into the desired single lasing line. Additionally, by intentionally engineering the periods of the introduce PT modulation, the lasing WGM mode can be systematically tuned and controlled. Although we only demonstrated lasing for two WGM orders, this mode selection concept is general and even in principle valid with arbitrarily spectral